# From Metrics to Improvement: A Lifecycle-Aware LLM Feedback Framework for Research Software Quality

Nafis Tanveer Islam[1†], Nafiseh Soveizi[1†], Yutong Li[1], Zhiming Zhao[1*]

[1]Multiscale Networked Systems (MNS), University of Amsterdam, Science Park, Amsterdam, NH, The Netherlands.

*Corresponding author(s). E-mail(s): z.zhao@uva.nl; Contributing authors: n.t.islam@uva.nl; n.soveizi@uva.nl; yutong.li4@student.uva.nl;
†These authors contributed equally to this work.

**Abstract**

Research software is increasingly central to scientific workflows, yet it is often developed by researchers with limited software engineering expertise. This can lead to quality issues that hinder maintainability, reproducibility, reuse, and sustainability. Existing static analysis tools can identify such issues, but their outputs often require expert interpretation and provide limited support for translating quality assessments into actionable improvements. To address this gap, we propose a lifecycle-aware framework that integrates quantitative software quality assessment with Large Language Model (LLM)-based code refinement. The framework comprises two stages. First, a lifecycle-aware Quality Model is developed from established software quality standards and practitioner requirements. The model defines five quality dimensions and 25 candidate metrics, of which 14 are operationalized using existing analysis tools and custom measurements. Second, the resulting quality diagnostics are used as structured feedback within an iterative LLM-based refinement process, enabling generated improvements to be repeatedly reassessed against the Quality Model. We evaluate the framework on notebook-centric research software using multiple LLMs and compare iterative structured feedback with single-step feedback and unstructured prompting. The results show improvements in specific quality attributes, particularly code duplication and structural quality, while also revealing trade-offs among maintainability, code size, documentation, and complexity. These findings demonstrate the potential of metric-driven LLM feedback for research software quality improvement while highlighting its inherently multi-objective nature .*

# 1 Introduction

Computing technologies have become fundamental to modern scientific research workflows, enabling large-scale data processing, complex simulations, and reproducible experimentation. Among these technologies, *software* plays a crucial role in wrapping workflows by automating tasks, analyzing data, and visualizing results across domains such as climate science, healthcare, and engineering Kanewala and Bieman (2014); World Meteorological Organization (2026). This type of software, commonly referred to as *research software*, encompasses a broad spectrum of artifacts developed by and for researchers, ranging from lightweight analysis scripts to large-scale scientific infrastructures Eisty et al. (2018). Such software supports computationally intensive tasks, including large-scale simulations and data-driven analysis, and is essential for solving real-world problems across scientific domains Islam and Zhao (2025a); Volentir et al. (2025); Islam and Zhao (2025b); Kelly and Sanders (2008).

Research software can be categorized into different tiers based on its scope and intended use Nenadic et al. (2025); Felderer et al. (2025). Tier 1 software typically consists of exploratory artifacts such as scripts or Jupyter Notebooks that address a single research question. Tier 2 includes reusable tools developed within research groups, while Tier 3 refers to community-level, production-grade systems requiring long-term maintenance and governance. Although Tier 1 software is often informal and short-lived, it forms the foundation upon which higher-tier systems are built. Consequently, its quality directly impacts the scalability, reusability, and sustainability of scientific software ecosystems.

Ensuring software quality is critical for open science, particularly for *reproducibility*, *reusability*, and *transparency*. Poor-quality research software can lead to irreproducible results, increased technical debt, and limited reuse across projects and environments Soveizi et al. (2023). Empirical studies have shown that many research artifacts fail to meet basic quality standards, especially in exploratory settings where rapid development is prioritized over maintainability Pimentel et al. (2019); Wang et al. (2020b). For instance, large-scale analyses of Jupyter Notebooks reveal widespread issues such as code duplication, lack of documentation, and inconsistent execution order, all of which hinder reproducibility and reuse Koenzen et al. (2020); Pimentel et al. (2021); Wang et al. (2021). Moreover, studies report that a majority of researchers prioritize obtaining results over writing clean and maintainable code, further exacerbating quality issues Kery et al. (2017).

*The source code and experimental data are publicly available at https://github.com/QCDIS/Software_Quality_Control_LLM

To systematically assess software quality, several *software quality models* have been proposed, such as ISO/IEC 25010, CISQ, and the SIG/TÜViT model, which define high-level quality dimensions including maintainability, reliability, and security ISO (2011); Consortium for IT Software Quality (CISQ) (2016). These models rely on measurable *quality metrics*, such as cyclomatic complexity or code duplication, to provide quantitative insights into software quality. While such models are well established in industrial contexts, they are not directly applicable to research software, particularly Tier 1 artifacts, which are often informal, rapidly evolving, and lack structured engineering practices. Recent initiatives, such as EVERSE EVERSE Project (2025), highlight the need for research-specific quality frameworks that account for the unique characteristics of scientific workflows.

Among Tier 1 research artifacts, Jupyter Notebooks have become a dominant tool in data science and AI due to their interactive and flexible nature, allowing researchers to combine code, data, and documentation in a single environment Kluyver et al. (2016); Beg et al. (2021). They are also central to Virtual Research Environments (VREs), supporting collaborative and reproducible workflows Pelouze et al. (2025); Wang et al. (2025). Notebooks are widely adopted not only for research but also for education and training purposes Johnson (2020). However, their exploratory nature often encourages poor software engineering practices, resulting in non-modular code, weak dependency management, and limited testing Wang et al. (2020a); Chattopadhyay et al. (2020); Grotov et al. (2022). As a result, despite their popularity and potential for reproducibility, notebooks frequently suffer from low software quality. At the same time, existing approaches for addressing these issues remain fragmented: static analysis tools can assess quality but require expertise to interpret PyCQA (2024); Lacchia (2025); Kucherenko (2024), while Large Language Models (LLMs) can assist with code generation and repair but often lack structured, metric-driven guidance Sallou et al. (2024); Sun et al. (2025); Patil (2025); Shi et al. (2025); Shen et al. (2025). Existing LLM-based tools also tend to focus on specific aspects, such as security, rather than holistic software quality improvement Nunez et al. (2024); Islam et al. (2024).

This motivates the need for approaches that jointly address *quality modeling with assessment* and *quality improvement* in a unified manner. Accordingly, this paper addresses the following question: *How can software quality in Tier 1 research artifacts be systematically measured and improved by integrating metric-based evaluation with LLM-driven refinement?* To answer this question, we propose a lifecycle-aware, feedback-driven framework that combines quantitative quality assessment with iterative LLM-based code refinement. In our proposed framework, software quality metrics act as structured signals that guide the software refinement process, enabling continuous and measurable improvement of research software quality.

The novelty of our work lies in integrating a lifecycle-aware quality model with an LLM-based feedback loop tailored to notebook-centric research workflows, bridging the gap between quantitative evaluation and automated improvement.

The main contributions of this paper are as follows:

- We define a lifecycle-aware set of software quality metrics tailored to Tier 1 research software, grounded in existing standards and adapted to notebook-centric workflows.
- We design a feedback-loop-based system that integrates static analysis metrics with LLM-driven code refinement to iteratively improve software quality.
- We evaluate the proposed approach through quantitative experiments and a qualitative case study, demonstrating its effectiveness in improving code quality.

The remainder of this paper is organized as follows. Section 2 reviews existing approaches to research software quality assessment and LLM-based quality improvement and identifies the key research gaps. Section 3 presents the overall methodology, including the requirement analysis, framework design, and research questions. Section 4 describes the development of the lifecycle-aware Quality Model, including metric selection, operationalization, and lifecycle-aware integration. Section 5 presents the LLM-based quality improvement framework, including its methodology, system design, and experimental evaluation. Section 6 discusses the findings in relation to the research questions and outlines the limitations of the proposed approach. Finally, Section 7 concludes the paper and presents directions for future work.

# 2 Related Work

In this section, we critically investigate all the approaches for analyzing and improving software quality, and finally, we analyze work on automatically evaluating software.

## 2.1 Quality Issues in Software

A significant amount of work has documented critical quality issues in Jupyter Notebooks. Wang et al. (2020b) analyzed a large corpus of public notebooks and found that many exhibited substandard code quality, including non-adherence to recommended coding practices, the presence of unused variables, and the use of deprecated functions, and noted a lack of dedicated quality-assurance tools explicitly tailored to notebooks. Koenzen et al. (2020) reported that approximately 1 in 13 cells in real-world notebooks are duplicated, undermining maintainability and reproducibility and increasing the risk of inconsistent behavior and technical debt Mohamed et al. (2026). Pimentel et al. (2021) examined reproducibility problems in real-world notebooks and concluded that reproducibility is still far from ideal. However, they emphasized that notebooks have significant potential to support reproducible research, provided that better coding practices and appropriate tooling are adopted. Rule et al. (2018) similarly observed that, while many researchers exploit the interactive capabilities of notebooks for exploratory experimentation, they often struggle to transform these exploratory artifacts into well-documented, reproducible analyses that others can readily understand and execute. Taken together, these studies Wang et al. (2020b); Koenzen et al. (2020); Pimentel et al. (2021); Rule et al. (2018) consistently indicate that Jupyter Notebooks, particularly when used in Tier-1 research settings, are prone to quality issues and call for dedicated methods and tools to address them. Recently, with the increased usage of LLMs with coding assistance, while research shows that it has increased development speed, it is still unclear how these generative models affect developers' cognitive offloading or ensure software quality Mohamed et al. (2026).

## 2.2 Approaches for Improving Software Quality

Approaches for improving software quality span both research-specific tools designed for notebook-centric workflows and more general software engineering methods that have recently been explored in research contexts. While static, rule-based approaches are often tailored to the characteristics and constraints of research software, many emerging learning-based methods originate from general software engineering settings and do not explicitly account for the specific requirements of research software, such as exploratory development, reproducibility, and continuously evolving artifacts. In this section, we distinguish between these two classes of approaches and discuss the implications of this mismatch for applying learning-based quality-improvement methods in research software.

### *Static and Rule-Based Tools*

A first line of work focuses on static analysis and rule-based techniques that are explicitly designed for research software, particularly notebook-centric workflows such as Jupyter Notebooks. Titov et al. (2022) propose an automated cell re-splitting algorithm that improves the logical structure of notebooks by decomposing long, monolithic cells, thereby enhancing modularity and reuse. Pimentel et al. (2021) introduce Julynter, a JupyterLab extension that provides linting and automated suggestions based on both language-agnostic and language-specific analyses, targeting issues related to notebook structure, execution history, and cell dependencies. Subotić et al. (2022) present NBLyzer, a static analysis framework that simulates alternative execution orders to detect potential correctness and reproducibility issues before execution. Quaranta et al. (2024) develop Pynblint, a Python-based static analyzer for Jupyter Notebooks that assesses adherence to coding standards and best practices and can be integrated into CI/CD pipelines.

While these tools effectively surface specific classes of quality issues—such as execution order violations, structural deficiencies, and coding-style inconsistencies—they are largely limited to isolated quality aspects and single development environments. Moreover, they require substantial manual effort from researchers to interpret diagnostics and apply fixes, and they do not provide integrated, lifecycle-aware support that continuously guides quality improvement as research software evolves.

### *LLM-Based Quality Improvement Approaches*

Recent studies have investigated the use of LLMs for improving different aspects of software quality. Katzy et al. (2025) examine LLM-generated code comments and show that comment correctness represents an important quality concern when LLMs are used for software development. Other approaches improve code generation by incorporating broader software context. For example, Zhang et al. (2023) introduce RepoCoder, a retrieval-augmented code completion framework that exploits repository-level information to generate more context-aware code, together with Repo-Eval, a benchmark for evaluating code completion at line, API-call, and function-body levels. Shrivastava et al. (2022) similarly investigate repository-level prompt generation to provide LLMs with relevant contextual information during code generation.

More recent approaches move beyond contextual prompting toward explicit quality assessment and iterative improvement. PELLI Krebs and Mazumdar (2026) integrates LLM-based software generation with iterative analysis and quantitative quality assessment, while CodeQUEST Lius et al. (2025) evaluates and improves generated code across multiple software quality dimensions. Arguello Ruiz et al. (2025) propose a four-step process comprising planning, execution, verification, and analysis and adjustment to support continuous software quality improvement with LLMs. Other studies have examined LLMs as complements or alternatives to conventional quality-assurance tools. AlOmar (2025), for example, compares LLM-based feedback with the Programming Mistake Detector (PMD) in programming education, while Umre et al. (2026) formalize software engineering metrics for evaluating the quality of LLM-generated code. Related work has also investigated LLM-based improvement of specific quality attributes, particularly software security Nunez et al. (2024); Islam et al. (2024).

These studies demonstrate that LLMs can support code assessment, repair, refactoring, and quality improvement, and that quantitative software quality measures can be incorporated into LLM-based evaluation and refinement. However, these approaches are primarily developed for general software engineering contexts rather than research software and therefore do not explicitly address the characteristics of Tier 1 notebook-centric development, such as exploratory programming, evolving artifacts, reproducibility, and reuse. Moreover, existing approaches do not integrate metric-driven LLM refinement with a lifecycle-aware Quality Model derived from research software standards and practitioner requirements. This leaves a gap in connecting research-software-specific quality requirements and lifecycle context with operationalized quality diagnostics that can systematically guide iterative LLM-based refinement.

## 2.3 Research Gaps

The literature reviewed above demonstrates significant progress in both notebook-oriented software quality assessment and LLM-assisted code improvement. Research-software quality tools such as Julynter, NBLyzer, and Pynblint provide mechanisms for detecting quality issues in Jupyter notebooks, but primarily focus on assessment and leave the interpretation and resolution of identified issues to researchers. In parallel, recent LLM-based approaches have progressed from contextual prompting toward metric-based evaluation and iterative code improvement. However, these approaches are predominantly designed for general software engineering contexts and do not explicitly account for the characteristics of Tier 1 research software, such as exploratory development, evolving artifacts, reproducibility requirements, and lifecycle-dependent quality concerns. Consequently, a gap remains in connecting research-software-specific quality assessment with automated, context-aware quality improvement.

This analysis reveals three research gaps.

First, there is limited integration between research-software-specific quality assessment and LLM-based quality improvement. Although recent LLM-based approaches incorporate software quality measures into code evaluation and refinement, they are not explicitly grounded in quality requirements tailored to notebook-centric research software. Conversely, existing notebook quality tools provide specialized assessments but do not systematically translate their diagnostics into structured feedback for automated refinement. A framework is therefore needed to connect measurable quality concerns specific to Tier 1 research software with LLM-based code improvement.

Second, existing approaches provide limited support for lifecycle-aware quality assessment and improvement. Quality priorities can change as research software progresses through different lifecycle stages, from Initialization and Implementation to Publication, Deployment & Platform Integration, and Community Feedback. However, existing notebook quality tools and LLM-based improvement approaches generally do not use lifecycle context to determine which quality concerns should be assessed or prioritized. Incorporating lifecycle information can enable quality assessment and refinement to better reflect the evolving needs of research software.

Third, there is limited traceability between research software quality requirements, measurable diagnostics, and LLM-generated improvements. Although existing approaches may employ quality metrics to evaluate or guide generated code, these metrics are not typically organized within a research-software-specific Quality Model that relates them to broader quality dimensions, practitioner requirements, and lifecycle context. Such traceability is important for explaining why a quality issue is assessed, how it relates to research software quality objectives, and how the resulting diagnostic should guide subsequent refinement.

These gaps motivate the methodology proposed in this paper. We first conduct a requirement analysis combining established software quality standards with practitioner perspectives to identify the functional and non-functional requirements of a quality improvement framework for Tier 1 research software. In Stage 1, these requirements are translated into a lifecycle-aware Quality Model and operationalized through measurable quality metrics. In Stage 2, the resulting diagnostics are used as structured feedback within an iterative LLM-based refinement process, connecting research-software-specific quality assessment with automated quality improvement in a closed feedback loop.

# 3 Methodology

The research gaps identified in Section 2 indicate that improving the quality of Tier 1 research software requires more than combining existing software quality assessment tools with LLMs. Before designing such a framework, it is necessary to identify the quality requirements for notebook-centric research software and its users' needs. These requirements will provide the design foundation for the proposed framework.

The proposed methodology is illustrated in Figure 1. It begins with a requirement analysis that combines evidence from established software quality standards and practitioner perspectives to derive the functional and non-functional requirements for a quality improvement framework targeting Jupyter notebook-centric research software.

These requirements provide the foundation for the proposed two-stage methodology. In Stage 1, we develop a lifecycle-aware Quality Model that provides structured and quantitative quality assessment. In Stage 2, the resulting quality metrics are used as structured feedback within an iterative LLM-driven refinement process, establishing a closed-loop framework for automated software quality improvement.

The remainder of this section proceeds as follows. We first present the requirement analysis, which derives the functional and non-functional requirements of the proposed framework from established software quality standards and practitioner perspectives. We then describe how these requirements are translated into the design of the proposed two-stage framework, comprising a lifecycle-aware Quality Model for quantitative software quality assessment and an LLM-based quality improvement framework. Finally, we formulate the research questions that guide the implementation and evaluation of the proposed framework.

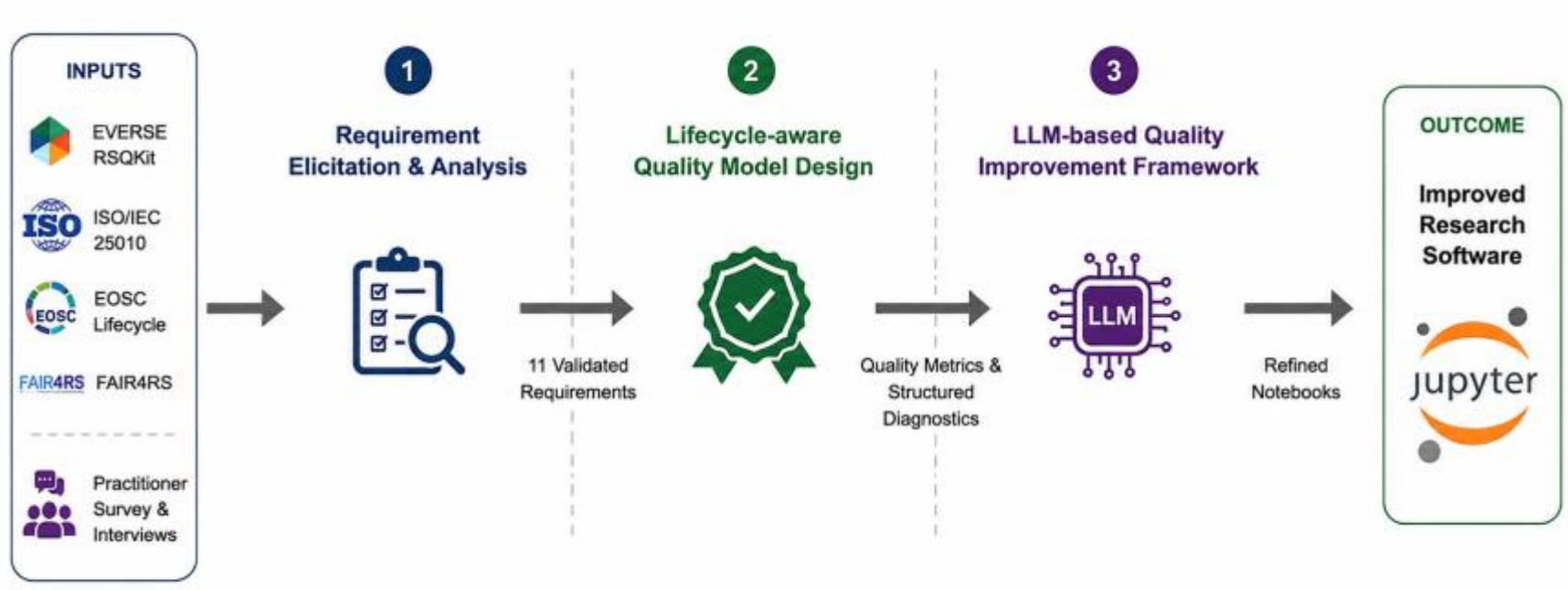

**Fig. 1**: Overview of the proposed methodology, from requirement analysis to Quality Model design and LLM-based quality improvement.

## 3.1 Requirement Analysis

The objective of the requirement analysis is to identify the capabilities that a software quality improvement framework should provide for Tier 1 notebook-centric research software. This analysis is needed because research software quality is difficult to address in a uniform way. Established software quality models, such as ISO/IEC 25010 ISO (2011) and CISQ Consortium for IT Software Quality (CISQ) (2016), provide well-defined quality dimensions and metrics, but they originate from different communities, including industrial software engineering, research software initiatives, and FAIR/open-science frameworks. Each of these communities emphasizes different aspects of quality. Moreover, these models are primarily designed for structured software systems and often assume mature development processes. In contrast, research software, particularly Tier 1 notebook-centric artifacts, is exploratory, rapidly evolving, and frequently developed without formal software engineering practices. This creates a gap between standardized quality frameworks and the practical realities of research workflows. Research communities themselves are also highly heterogeneous. Researchers work across different scientific domains, follow diverse development practices, and prioritize different quality concerns depending on their goals, constraints, and level of software maturity. As a result, defining a unified and applicable notion of software quality for Tier 1 research software requires both standards-based grounding and practitioner validation.

To address this, we combine complementary evidence from established software quality standards and practitioner experience. Four established frameworks were considered: ISO/IEC 25010 ISO (2011), EVERSE RSQKit EVERSE Project Team (2024), FAIR4RS Netherlands eScience Center and DANS (2025), and the EOSC EOSC Task Force on Infrastructure for Quality Research Software (2026) research software lifecycle. ISO/IEC 25010 provides a general taxonomy of software quality characteristics, EVERSE RSQKit introduces research-software-specific quality concerns, FAIR4RS contributes principles related to findability, accessibility, interoperability, and reuse, and the EOSC research software lifecycle provides a temporal perspective by relating software quality concerns to different stages of research software development and reuse.

### *Survey Analysis*

To complement these standards with practitioner perspectives, we conducted a targeted survey involving 20 participants from scientific research and research-software environments. The participants represented a diverse group of domain researchers, particularly from biology, ecology, and environmental science, together with research software engineers, data and AI practitioners, students, and other academic roles. Participants had varying levels of programming experience and familiarity with Jupyter Notebook workflows, ranging from frequent notebook users to participants who used notebooks less regularly. Table 1 summarizes the available participant characteristics.

**Table 1**: Background characteristics of survey participants ($n = 20$).

| Characteristic | Category | % |
|---|---|---|
| Programming experience | 1-5 years | 15.0 |
| | 5-10 years | 30.0 |
| | >10 years | 55.0 |
| Jupyter Notebook use | Daily | 25.0 |
| | Weekly | 35.0 |
| | Monthly | 30.0 |
| | Rarely | 10.0 |

The survey examined four aspects of notebook-driven research software practice: (i) challenges encountered when reusing notebooks developed by others, (ii) self-reported coding and documentation practices, (iii) awareness of research software lifecycle stages, and (iv) prioritization of quality dimensions such as maintainability, FAIRness, openness, and sustainability. Across the responses, recurring concerns emerged around insufficient documentation, weak dependency management, poor modularization, and lack of structured testing. At the same time, participants reported uneven familiarity with lifecycle thinking and heterogeneous views on which quality dimensions matter most. All of our survey contents (Queries and Responses) and code to reproduce our methodology are available here[1].

Together, these observations highlight three core challenges that must be reflected in the framework requirements: *Community Diversity*, *Lifecycle Diversity*, and *Requirement Diversity*. We discuss these challenges below and use them to derive the functional and non-functional requirements of the proposed framework.

### *Challenge 1: Community Diversity.*

The first challenge arises from the diversity of research-software users and the quality problems they encounter when reusing notebooks developed by others. Our participants represented different scientific and technical backgrounds and had varying levels of programming experience and familiarity with Jupyter Notebook workflows. Despite this diversity, several recurring quality concerns emerged from the survey. As shown in Table 2, the most frequently reported issues were insufficient explanation or comments (90%), dependency or library problems (75%), poor modular structure (65%), and missing or failing tests (65%). Disorganized cell logic and difficulty understanding the reasoning behind code were each reported by 50% of participants.

These findings indicate that researchers encounter a combination of documentation, maintainability, reproducibility, and structural problems when reusing notebook-based research software. Other concerns, including code complexity, duplication, licensing, and security, were reported less frequently but remain relevant to software quality. The variation in both user backgrounds and reported concerns shows that quality assessment for Tier 1 research software needs to accommodate users with different levels of software-engineering expertise and address multiple aspects of quality rather than relying on a single criterion. This motivates requirements for multidimensional assessment and interpretable feedback that can be understood and acted upon by both domain researchers and software-oriented practitioners.

The survey was used to identify and prioritize practical quality concerns, rather than to define the metrics used to assess them. To connect these practitioner-reported concerns to measurable properties of research software, each concern was mapped to an associated quality metric identified through the standards- and literature-based analysis described above. The resulting mapping is shown in Table 2. The EOSC research software lifecycle EOSC Task Force on Infrastructure for Quality Research Software (2026) is subsequently used to determine the relevance of these metrics at different stages of the research software lifecycle. In this way, practitioner evidence establishes the practical relevance of the quality concerns, while the standards and literature provide the basis for their operational assessment.

[1] https://github.com/QCDIS/Software_Quality_Control_LLM

**Table 2**: Quality issues reported when reusing Jupyter Notebooks and their associated quality metrics.

| Issue | % | Associated Metric |
|---|---|---|
| Too little explanation or comments | 90% | Comment Density |
| Dependency or library issues | 75% | Dependency Management |
| Poor modular structure | 65% | Modularity |
| No or failing tests | 65% | Test Success Rate |
| Disorganized cell logic | 50% | Cohesion |
| Hard to understand reasoning behind code | 50% | Requirement Traceability |
| Incomplete or missing documentation | 40% | Documentation Quality |
| Excessive lines of code | 40% | Software Size (LoC) |
| Use of bad practices or code smells | 25% | Code Smells |
| High complexity / hard-to-follow logic | 25% | Cognitive Complexity |
| Lack of reproducibility | 25% | Code Reproducibility |
| Use of duplicated code | 25% | Code Duplication |
| Missing license / unclear reuse terms | 25% | Presence of License |
| Security issues / secrets in code | 10% | No Leaked Private Credentials |

***Challenge 2: Lifecycle Diversity.***

The second challenge concerns differences in how researchers engage with and perceive the research software lifecycle. Notebook-based research software is often developed iteratively, with activities ranging from initial exploration and implementation to publication, deployment, and subsequent reuse. However, researchers may participate in different parts of this process and may not explicitly conceptualize their work in terms of a software lifecycle.

To provide a structured basis for examining these differences, we use the research software lifecycle proposed by EOSC Task Force on Infrastructure for Quality Research Software (2026). The EOSC lifecycle reflects the iterative and evolving nature of research software and provides a common structure for relating quality concerns to different stages of research software development and reuse. The lifecycle stages were therefore defined from this established model rather than derived from the survey; the survey was used to examine participants' engagement with these stages.

Participants were asked which lifecycle stages they frequently encounter in their work. As shown in Table 3, Initialization, Implementation, and Publication were each reported by 40% of participants, while Planning was reported by 25%. Deployment and Platform Integration and Community Feedback were less frequently encountered, each being reported by 10% of participants. Importantly, 40% indicated that they do not explicitly think about their work in lifecycle terms. Since participants could report multiple stages, the percentages are not expected to sum to 100%.

These results indicate that research-software practitioners engage unevenly with different lifecycle stages and that explicit lifecycle awareness cannot be assumed. Consequently, applying the same set of quality criteria uniformly throughout a project's evolution may provide feedback that is irrelevant to the user's current activities. This motivates a lifecycle-aware assessment in which quality metrics and feedback can be adapted to the stage selected by the user, while keeping the interaction lightweight enough for researchers who do not routinely use formal software-engineering lifecycle concepts.

**Table 3**: Research software lifecycle stages frequently encountered by survey participants. Multiple selections were permitted.

| Lifecycle Stage | % |
|---|---|
| Initialization (e.g., developing research ideas) | 40% |
| Planning (e.g., defining goals and preparing the environment) | 25% |
| Implementation (e.g., coding, data analysis, and modelling) | 40% |
| Publication (e.g., preparing notebooks for reuse or publication) | 40% |
| Deployment & Platform Integration | 10% |
| Community Feedback | 10% |
| Do not explicitly think in lifecycle terms | 40% |

**Table 4**: Participant ratings of research software quality dimensions.

| Dimension | Slightly | Moderately | Very | Extremely |
|---|---|---|---|---|
| Technical Performance | 15% | 10% | 35% | 40% |
| FAIRness | 15% | 35% | 25% | 25% |
| Openness | 35% | 15% | 35% | 15% |
| Sustainability | 35% | 15% | 35% | 15% |

### *Challenge 3: Requirement Diversity.*

The third challenge concerns differences in how researchers prioritize software quality requirements. Research software quality is multidimensional, and the relevance of individual quality dimensions may vary depending on the research context, intended use, and stage of development. To examine these differences, participants were asked to rate the importance of four quality dimensions considered in our standards-based analysis: Technical Performance, FAIRness, Openness, and Sustainability. These dimensions were therefore not derived from the survey; rather, the survey was used to examine how practitioners perceive and prioritize them in practice.

As shown in Table 4, Technical Performance received the strongest ratings, with 75% of participants considering it very or extremely important. FAIRness showed a more moderate distribution, although 50% still rated it as very or extremely important. Opinions on Openness and Sustainability were more heterogeneous: for both dimensions, 50% rated them as very or extremely important, while 35% considered them only slightly important. These differences indicate that practitioners do not assign the same priority to all aspects of research software quality.

The results reinforce the need for a multidimensional quality model rather than a single aggregate notion of software quality. At the same time, the variation in priorities suggests that quality assessment should preserve the individual dimensions and provide feedback at an appropriate level of granularity, allowing users to understand which aspects of their software require attention. In combination with the standards-based analysis, these practitioner perspectives therefore motivate a framework that covers both immediate technical concerns and broader research-software qualities such as FAIRness, openness, and sustainability without assuming that all dimensions have equal relevance in every research context.

## 3.2 Framework Requirements

Overall, the survey shows that researchers encounter diverse quality challenges when reusing notebook-based software, engage unevenly with different lifecycle stages, and assign different priorities to software quality dimensions. These findings imply that a useful solution must do more than measure code quality: it must account for lifecycle context, support diverse quality requirements, remain accessible to users with different levels of software-engineering expertise, and transform quality assessments into actionable guidance.

Based on the combined evidence from the standards analysis and survey, we derived a set of functional and non-functional requirements for the proposed framework, following the ISO/IEC/IEEE 29148 principles. Only requirements that were relevant to the identified practitioner needs, traceable to the analyzed standards, and feasible within notebook-centric environments were retained. As an additional validation step, the derived requirements were reviewed with the survey participants through online interviews. Their feedback was used to confirm the relevance of the requirements to the identified practical challenges and to refine them before finalizing the framework. The final set comprises five functional requirements and six non-functional requirements, summarized in Table 5. The functional requirements emphasize lifecycle-aware assessment, seamless notebook integration, unified metric execution and reporting, standards-based natural-language feedback, and results presented at multiple levels of granularity. The non-functional requirements ensure that the framework remains lightweight, efficient, portable, extensible, privacy-preserving, and robust in practical research environments. Together, these validated requirements provide the foundation for the quantitative Quality Model developed in Stage 1 and the structured-diagnostics feedback loop employed in Stage 2.

**Table 5**: Validated requirements (5 Functional + 6 Non-Functional)

| Requirement | Description |
|---|---|
| **FR1** | FR1 Lifecycle-aware assessment selectable by lifecycle stage. |
| **FR2** | In-notebook initiation and flexible project input (file/folder). |
| **FR3** | Unified metric execution and reporting (score + interpretation). |
| **FR4** | Natural-language feedback linked to standards and best practices. |
| **FR5** | Multi-level result display (summary, project, file). |
| **NFR1–NFR6** | Lightweight setup, acceptable runtime, multi-platform operation, extensibility, privacy preservation, and robustness to malformed input. |

## 3.3 Framework Design

The Framework Requirements presented in the previous subsection define the functional and non-functional capabilities the proposed framework must meet. Rather than addressing these requirements through independent components, we organize them into a two-stage architecture that separates quality assessment from quality improvement while maintaining a continuous feedback loop between them.

The extracted requirements naturally fall into two complementary groups. Requirements related to lifecycle awareness, quantitative quality assessment, and unified reporting motivate the design of Stage 1, whose objective is to provide a consistent and interpretable assessment of Jupyter notebook-centric research software. Requirements concerned with actionable feedback and automated quality improvement motivate Stage 2, which uses the assessment results generated by Stage 1 to iteratively refine the notebook. Organizing the framework in this way allows quantitative quality assessment to directly drive automated software improvement while preserving a clear separation between evaluation and refinement.

Figure 2 illustrates how the two stages interact with the researcher throughout the research software lifecycle. During different lifecycle stages, the researcher submits a notebook to the lifecycle-aware Quality Model (Stage 1), which evaluates its quality and generates structured diagnostics according to the relevant lifecycle context. These diagnostics are then passed to the LLM-based enhancement component (Stage 2), where they are used as explicit guidance for code refinement. The refined notebook is returned to the researcher and can subsequently be reassessed as the software progresses through the lifecycle, establishing a continuous assessment-improvement cycle.

The detailed design of the lifecycle-aware Quality Model is presented in Section 4, while the LLM-based quality improvement framework is described in Section 5.

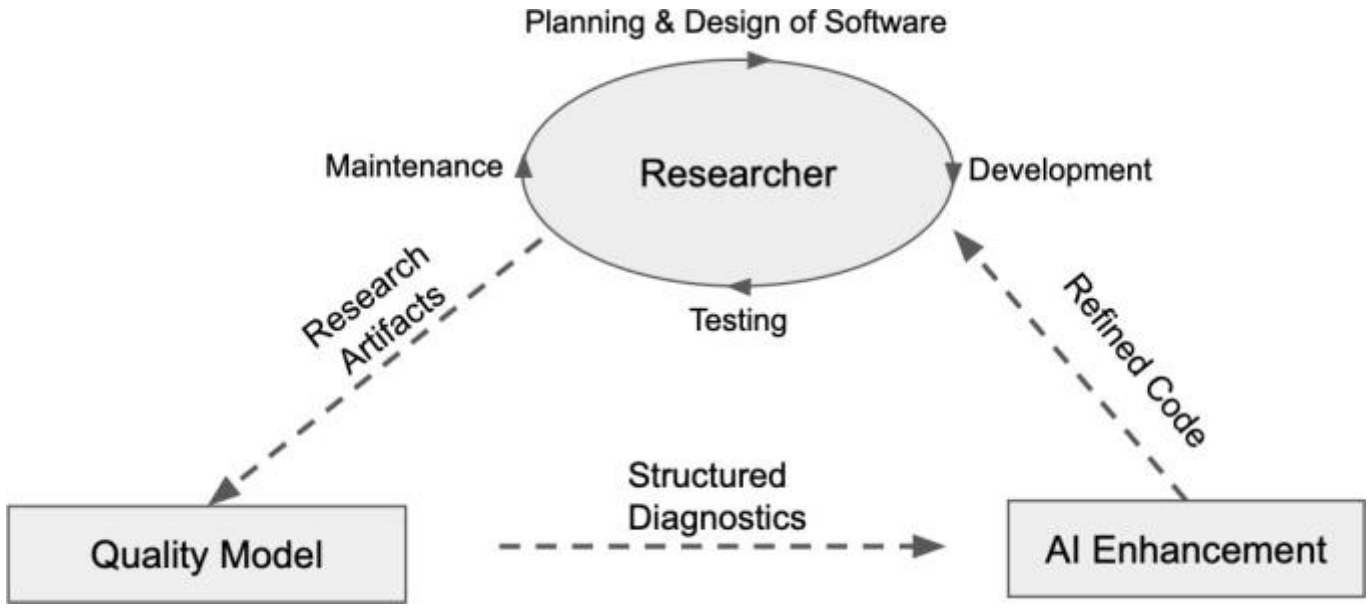

**Fig. 2**: Interaction model linking the researcher, the Quality Model, and the AI-agent across the research-software lifecycle.

## 3.4 Research Questions

Guided by the proposed methodology, we formulate the following research questions to structure the design and evaluation of the framework.

**RQ1:** How can a lifecycle-aware Quality Model be designed to systematically assess the quality of Tier 1 notebook-centric research software?

**RQ2:** How can the quality metrics produced by the Quality Model be used as structured feedback to guide iterative LLM-based refinement of research software?

**RQ3:** To what extent does the proposed feedback-driven framework improve the quality of Tier 1 research software in practice?

# 4 Stage One: Lifecycle-aware Quality Model

Building upon the validated requirements derived in Section 3, Stage 1 develops the lifecycle-aware Quality Model that provides the quantitative foundation of the proposed framework. The objective of this stage is to transform the identified requirements into a measurable quality assessment model for Jupyter notebook-centric research software by defining quality dimensions, evaluating the operational feasibility of candidate metrics, and integrating a selected subset with lifecycle information to generate structured quality diagnostics.

The development of the Quality Model comprises three sequential activities. We first translate the validated requirements into quality dimensions and candidate metrics. We then evaluate the operational feasibility of these metrics by mapping them to available assessment approaches. Finally, a selected subset of feasible metrics is implemented and integrated with lifecycle information to construct the executable lifecycle-aware Quality Model.

## 4.1 Quality Model Design

The first activity in constructing the lifecycle-aware Quality Model translates the validated requirements from Section 3 into measurable quality dimensions and candidate metrics. Five complementary dimensions were identified for notebook-centric Tier-1 research software: *Maintainability*, *Security*, *FAIRness*, *Functional Suitability*, and *Sustainability*. Together, they address software structure, correctness, security, reuse, and long-term sustainability. To ensure traceability, the dimensions were linked to the standards and practitioner evidence that motivated their inclusion. As summarized in Table 6, they are grounded in ISO/IEC 25010, FAIR4RS, EVERSE RSQKit, and the practitioner survey. Candidate metrics were then selected from existing software quality studies and research software quality frameworks based on their conceptual relevance to each dimension. At this stage, implementation feasibility was not used as a selection criterion; therefore, the candidate set includes both automatically measurable metrics and metrics requiring manual or indirect assessment.

The resulting conceptual Quality Model comprises five dimensions and twenty-five candidate metrics. Maintainability contains the largest number of metrics, reflecting the recurring concerns identified in both the standards review and practitioner survey regarding code organization, readability, complexity, and reuse. The other dimensions extend the model to secure reuse, FAIR software practices, functional correctness and verification, and longer-term dependency and sustainability concerns. The next subsection evaluates the operational feasibility of these candidate metrics and identifies those for which reliable measurement approaches are currently available.

## 4.2 Metric Operational Feasibility

The candidate Quality Model presented in the previous subsection provides a comprehensive conceptual representation of research software quality. However, not all candidate metrics can be measured reliably within a lightweight notebook-centric assessment workflow. Therefore, this activity evaluates the operational feasibility of the candidate metrics and identifies those for which suitable measurement approaches are currently available. To assess operational feasibility, candidate metrics were evaluated according to three criteria: (i) availability and maintenance status of existing tools, (ii) compatibility with notebook-centric research software, and (iii) reliability of metric computation. The evaluation considered widely used static analysis tools, including Pylint, Radon, JSCPD, Bandit, Gitleaks, and HowFairIs, together with a set of custom scripts developed to support notebook-specific quality measurements. Table 7 summarizes the operational feasibility of the candidate metrics, while Figure 3 provides an overview of the resulting feasibility categories. Among the twenty-five candidate metrics, eleven can be measured automatically using existing tools, three are partially supported, six require manual assessment or indirect evaluation, four currently lack a suitable measurement approach, and one requires direct user feedback.

**Table 6**: Candidate Quality Dimensions and Associated Metrics

| Quality Dimension | Source | # Metrics | Candidate Metrics |
|---|---|---|---|
| Maintainability | ISO/IEC 25010, RSQKit, Survey | 11 | Modularity, Cohesion, Architectural Complexity, Code Smells, Maintainability Index, Cognitive Complexity, Cyclomatic Complexity, Code Duplication, Technical Debt, Comment Density, Software Size (LoC) |
| Security | ISO/IEC 25010, RSQKit, Survey | 2 | Security Vulnerabilities, No Leaked Credentials |
| FAIRness | FAIR4RS, RSQKit, Survey | 4 | Presence of License, Public Repository, Rich Metadata, Documentation Quality |
| Functional Suitability | ISO/IEC 25010, RSQKit, Survey | 6 | Requirement Traceability, Unit Tests, Test Success Rate, Code Reproducibility, Defect Rate, Percentage of Assertions |
| Sustainability | ISO/IEC 25010, FAIR4RS, RSQKit, Survey | 2 | Dependency Management, User Satisfaction |

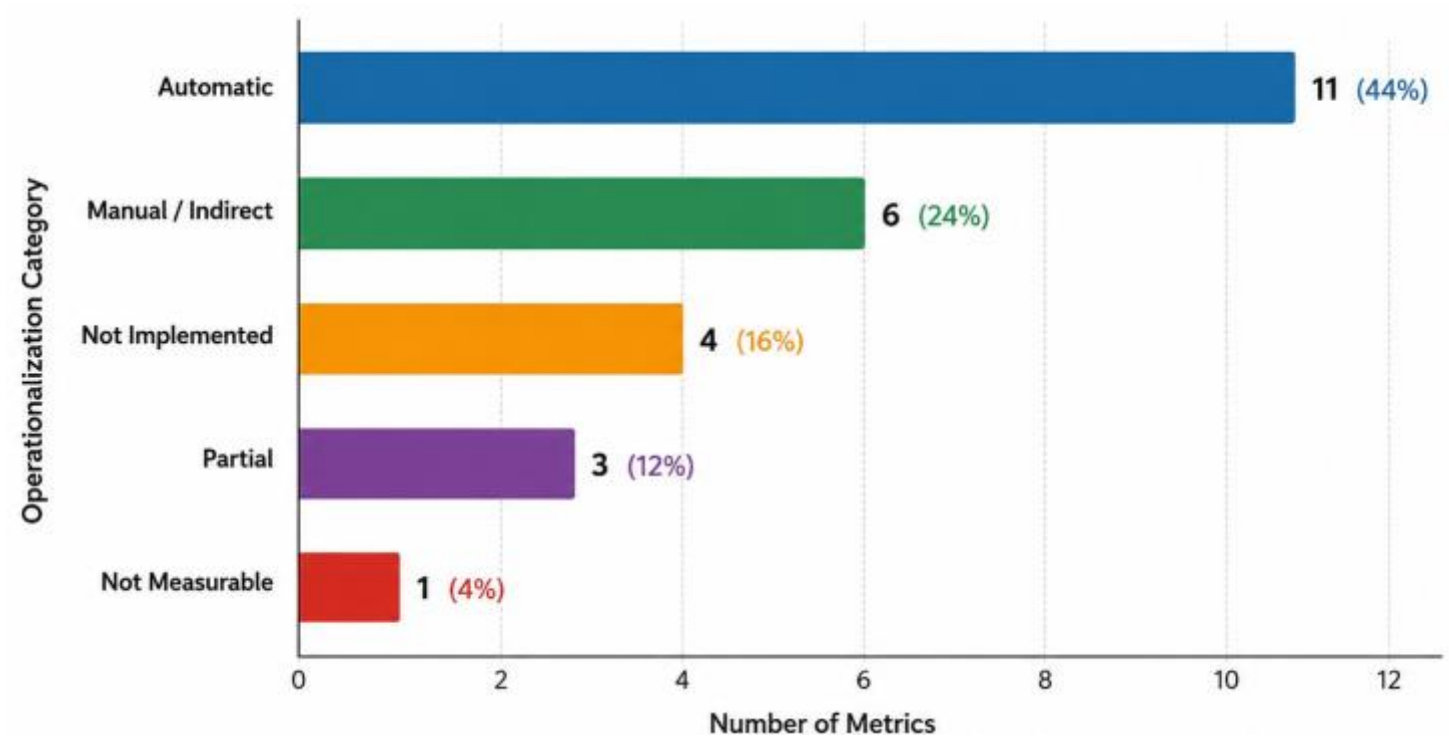

**Fig. 3**: Distribution of the 25 candidate metrics across operationalization categories.

**Table 7**: Distribution of the 25 candidate metrics across operational feasibility categories.

| Metric | Tool | Status |
|---|---|---|
| Code Smells | Pylint | Automatic |
| Maintainability Index | Radon | Automatic |
| Cyclomatic Complexity | Radon | Automatic |
| Code Duplication | JSCPD | Automatic |
| Comment Density | Radon / Custom Script | Automatic |
| Software Size (LoC) | Custom Script | Automatic |
| Percentage of Assertions | Custom Script | Automatic |
| Presence of License | HowFairIs | Automatic |
| Public Repository | HowFairIs | Automatic |
| No Leaked Credentials | Gitleaks | Automatic |
| Security Vulnerabilities | Bandit | Automatic |
| Dependency Management | Custom Script | Partial |
| Rich Metadata | HowFairIs | Partial |
| Documentation Quality | HowFairIs | Partial |
| Modularity | Manual Review | Manual |
| Cohesion | Manual Review | Manual |
| Architectural Complexity | Manual Review | Manual |
| Requirement Traceability | Manual Review | Manual |
| Cognitive Complexity | Manual Review | Manual |
| Technical Debt | Indirect Assessment | Manual / Indirect |
| Unit Tests | Not Available | Not Currently Operationalizable |
| Test Success Rate | Not Available | Not Currently Operationalizable |
| Code Reproducibility | Not Available | Not Currently Operationalizable |
| Defect Rate | Not Available | Not Currently Operationalizable |
| User Satisfaction | User Feedback | Not Measurable |

The results demonstrate that more than half of the candidate metrics can be measured automatically or semi-automatically using existing tooling. Metrics requiring manual assessment generally depend on human judgment or project-specific knowledge, whereas metrics that are not currently operationalizable rely on testing infrastructure, reproducibility environments, or defect-tracking information that is not consistently available across notebook-centric research projects. For the current implementation, we selected six maintainability-related metrics: Code Smells, Maintainability Index, Code Duplication, Comment Density, Software Size (LoC), and Cyclomatic Complexity. These metrics can be computed directly and repeatedly from notebook source code using lightweight static-analysis tools and provide complementary evidence about code quality, maintainability, redundancy, documentation, size, and structural complexity. This makes them particularly suitable for generating actionable feedback and evaluating changes during iterative LLM-based code refinement. The remaining candidate metrics are retained in the conceptual Quality Model and provide directions for extending the implementation to additional quality dimensions in future work.

## 4.3 Lifecycle-Aware Model Integration

The final activity integrates the six implemented metrics with lifecycle information and a common diagnostic representation to support consistent, lifecycle-aware quality assessment.

A key challenge in integrating multiple quality assessment tools is that they produce outputs with different formats, scales, and interpretations. To address this challenge, the outputs of the implemented metrics are transformed into a common diagnostic representation comprising the metric name, measured value, assessment status, interpretation, associated quality dimension, and relevant lifecycle stage. This unified representation enables heterogeneous tool outputs to be processed consistently and presented through a common reporting interface. The current executable model implements the six maintainability-related metrics selected in the previous activity. This focused implementation keeps the assessment lightweight, repeatable, and scalable while maintaining traceability to the broader conceptual Quality Model.

**Table 8**: Maintainability-related metrics implemented in the current framework

| Metric | Tool | Assessment Role |
|---|---|---|
| Code Smells | Pylint | Code-quality issues |
| Maintainability Index | Radon | Overall maintainability |
| Cyclomatic Complexity | Radon | Structural complexity |
| Code Duplication | JSCPD | Code redundancy |
| Comment Density | Radon / Custom Script | Code documentation |
| Software Size (LoC) | Custom Script | Software size |

By combining the implemented maintainability-related metrics with lifecycle context and a unified diagnostic representation, the resulting model provides structured and actionable quality diagnostics. These diagnostics form the basis for the LLM-based quality improvement process presented in the next section.

# 5 Stage Two: Using LLM to Enhance Software Quality

Stage 2 converts the structured diagnostics produced in Stage 1 (Section 4) into structured feedback that guides LLM-based code refinement and aims to improve software quality. Our objective in improving software quality is to help suggest higher-quality code, and researchers will ultimately make the final call on using the software. This helps researchers across different domains adapt key software quality measures such as maintainability, scalability, and security. Similar to the previous section, we also divide Stage 2 into three components, namely i) Methodology, ii) System Design, and finally iii) Experimental Analysis.

## 5.1 Methodology

To improve research software quality, we use the diagnostic report to guide the LLM-based refinement process through the following steps:

**Data Ingestion.** The ingestion layer processes each project by iterating through its file structure, extracting the relevant source files, and attaching all corresponding Stage 1 diagnostics to each code unit. Each code file is therefore paired with its metric results, lifecycle stage, and associated diagnostic messages, forming a coherent evidence package for the refinement process. This unified representation ensures consistent handling of project inputs and enables lifecycle-aware reasoning throughout Stage 2.

**Prompting Design.** We develop a prompting architecture that transforms file-level diagnostics and lifecycle tags into compact, lifecycle-aware evidence bundles. These bundles integrate metric violations, code excerpts, contextual metadata, and applicable quality guidelines into a constrained and reproducible prompt format. The design emphasizes semantic preservation and alignment with software quality dimensions, ensuring the LLM focuses on the most relevant weaknesses while maintaining compatibility with code stylistic conventions.

**LLM Inference & Program Repair.** The LLM agent uses the evidence bundles to produce refined versions of the affected code units. The model generates revised files that enhance documentation, reduce duplication, improve modularity, and address the specific quality issues identified in Stage 1. The refinement process is lifecycle-aware: the AI agent is guided to produce changes appropriate for the current lifecycle stage, emphasizing the quality concerns relevant to that stage. Each modification is traceable to specific metrics, allowing the refined artifacts to be evaluated directly against the Quality Model established in Stage 1.

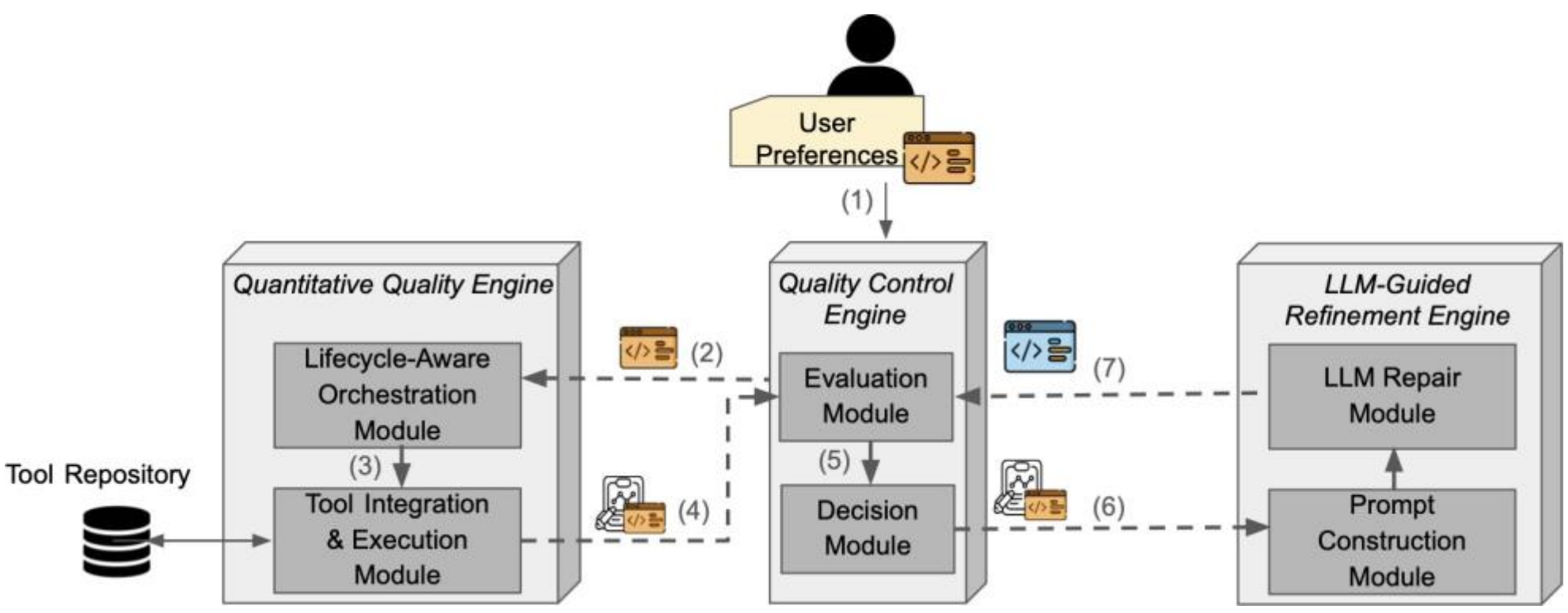

**Fig. 4**: Overview of our proposed System

## 5.2 System Design

Building on the methodology, the system integrates the outputs of Stage 1 (a lifecycle-aware Quality Model) and Stage 2 (an AI-guided refinement process) into a closed, iterative improvement loop. These stages define the approach's analytical and refinement capabilities.

We introduce a coordinating layer, *the Quality Control Engine*, which governs the interaction between quantitative assessment and the LLM-Guided Refinement Engine. This engine manages information flow, applies user-defined preferences and thresholds, and orchestrates the verify–rescore loop. By positioning the Quality Control Engine between the Quantitative Quality Engine and the AI-Guided Refinement Engine, the system ensures that refinements are lifecycle-aware, threshold-driven, and systematically validated, as illustrated in Figure 4. Therefore, we introduce the *Quality Control Engine*, which serves as the central decision-making layer governing the interaction between quantitative assessment and AI-guided refinement. It (i) interprets the metric values produced in Stage 1 as a unified quality score, (ii) decides when LLM-based repair is required, and (iii) governs the iterative verify-rescore loop. By positioning the Quality Control Engine between the Quantitative Quality Engine and the AI-Guided Refinement Engine, the system ensures that refinements are lifecycle-aware, threshold-driven, and systematically validated, as illustrated in Figure 4.

As shown in Figure 4, the system is composed of three cooperating modules: (a) the *Quality Control Engine* (Stage 1), which aggregates results, applies user preferences, and governs decision-making; (b) the *Quantitative Quality Engine* (Stage 2), which performs lifecycle-aware static analysis and produces structured diagnostics; and (c) the *LLM-Guided Refinement Engine* (Stage 2), which generates targeted code refinements when quality thresholds are not met. Together, these modules form a closed evaluate–decide–refine loop.

### 5.2.1 Quality Control Engine

The Quality Control Engine serves as the central coordinating component. It receives project files and user preferences, including metrics. It then manages evaluation and decision-making through two internal modules:

- **Evaluation Module.** This module requests a quantitative assessment from the Quantitative Quality Engine (Step 2). After receiving the diagnostic report from the six tools we used (Pylint, Radon, JSCPD, Bandit, Gitleaks, and Howfairis), it applies the user-defined metric weights and computes the final aggregated quality score.
- **Decision Module.** This module interprets the aggregated diagnostic report produced by the Evaluation Module. The results from the Decision Module are forwarded, along with the source code, to the LLM-Guided Refinement Engine. Refined outputs are returned to the Evaluation Module, which initiates a new assessment cycle, thereby sustaining the iterative verify–rescore loop.

### 5.2.2 Quantitative Quality Engine

The Quantitative Quality Engine executes the static-analysis pipeline defined in Stage 1. It receives an analysis request from the Evaluation Module and returns a structured diagnostic report. Internally, it consists of two modules:

- **Lifecycle-Aware Orchestration Module**.This module selects the metrics relevant to the current phase of the research software lifecycle and constructs an analysis plan accordingly.
- **Tool Integration & Execution Module.** This module executes the selected metrics using external static-analysis tools (e.g., Bandit, Gitleaks, Howfairis, JSCPD, Radon) and custom scripts. Results are parsed, normalized, and returned to the Evaluation Module as a unified diagnostic report.

### 5.2.3 LLM-Guided Refinement Engine

The LLM-Guided Refinement Engine receives source code and associated diagnostics, generates targeted refinements, and returns revised artifacts to the Evaluation Module for re-assessment. We divide the engine into two components:

- **Prompt Construction Module.** This module transforms structured diagnostics into a compact evidence package comprising violated metrics, code excerpts, lifecycle context, and applicable quality guidelines.
- **LLM Repair Module.** Guided by the evidence package, the language model generates refined versions of the affected code units. These refinements are returned to the Evaluation Module (Step 7), closing the verify–rescore loop.

## 5.3 Experimental Analysis

To evaluate the quality of LLM-generated code, we applied the operationalized quality metrics defined in Section 4 and summarized in Table 8. While we extracted 277 repositories, we randomly selected 10 for the feedback-based experiments (RQ2) because experimenting with LLMs is highly expensive, with cost increasing linearly with token count. We primarily experimented with OpenAI's paid LLMs and subsequently evaluated four local LLMs from HuggingFace. To run the LLMs locally, we used a single NVIDIA L40S with 46 GB of VRAM.

## 5.4 Dataset Preparation

We used the ENVRI Jupyter Notebook Search Interface to find a suitable collection of scientific projects ENVRI Community (2025). This search engine indexes publicly accessible Jupyter Notebooks from environmental and Earth science domains. We picked areas relevant to ecological and Earth science research, so we used the keywords ”ocean” and ”forest” in our query. The search results linked directly to GitHub projects, so we could download these projects using custom Python scripts for batch quality assessment. We developed a Python script that automates the download of public Git repositories containing Jupyter Notebooks. In total, we collected 277 research projects. Many of these projects contained multiple notebooks, so we ended up with a dataset of 2,796 Jupyter Notebook files. All downloaded repositories were saved in a directory and are available on our GitHub.

### 5.4.1 Characterizing Notebook Software Quality

We first analyze the 277 repositories we collected to understand the baseline quality of notebook-centric research software.

***Software Size (LoC)***

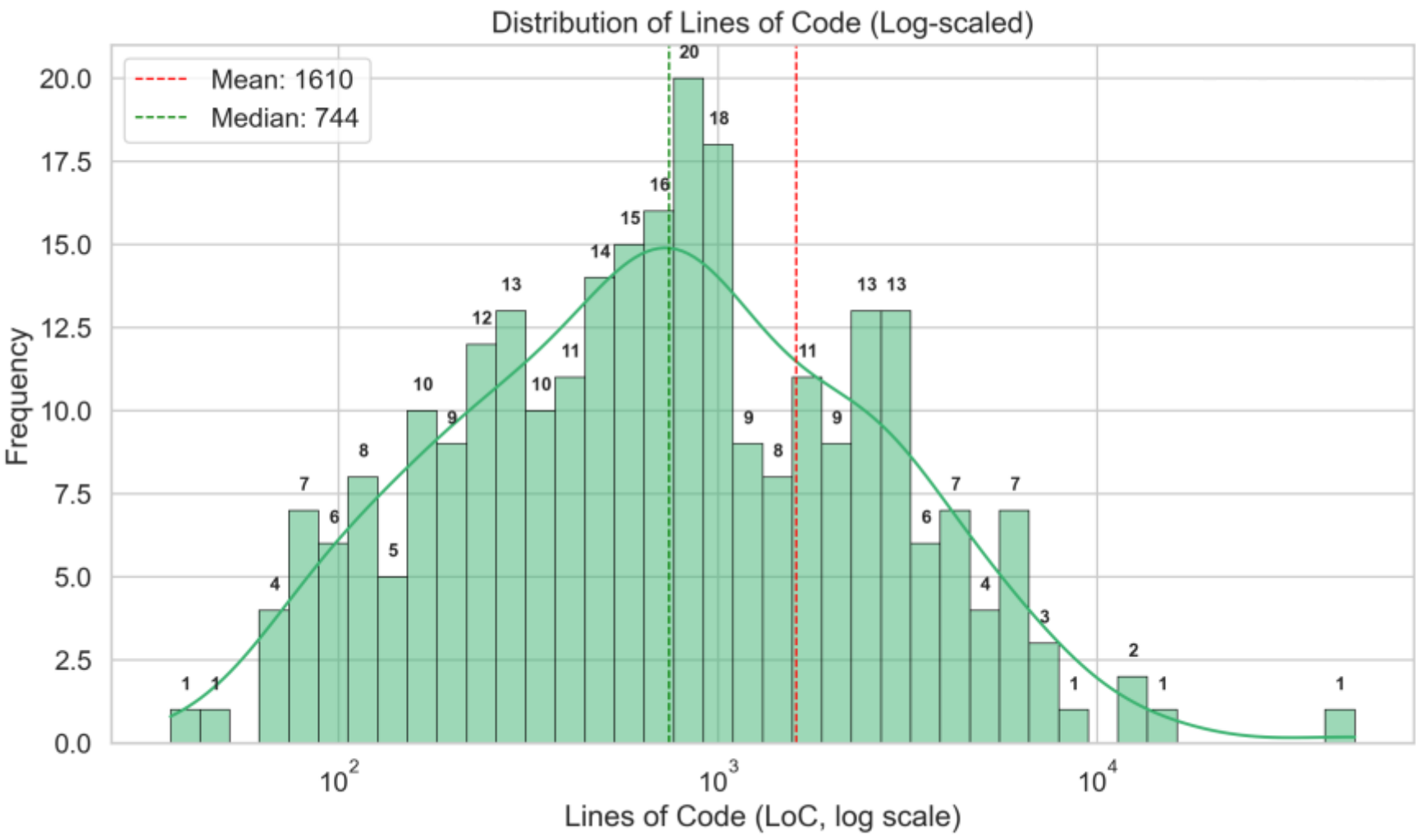

**Fig. 5**: Distribution of Lines of Code (Log-scaled)

From the log-scaled histogram in Figure 5, we see that the majority of software projects are in a moderate range for code size. There are some significant outliers, but most notebooks are clustered around 744–1,610 lines of code (median and mean, respectively). This suggests that many research notebooks are of a reasonable size and not excessively long or complex.

## Maintainability Index (MI)

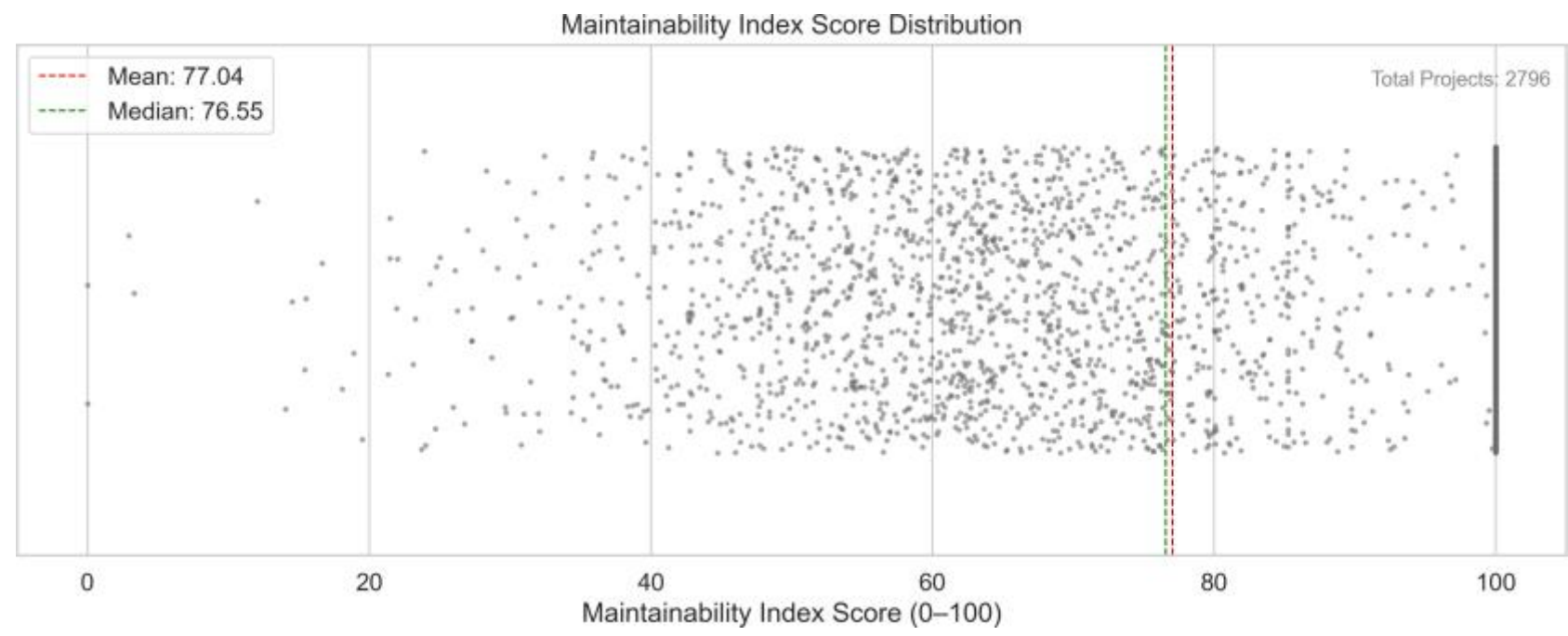

**Fig. 6**: Maintainability Index Score Distribution

From the strip plot in Figure 6, we observe that the maintainability scores are mainly concentrated toward the higher end. The mean is 77.04, and the median is 76.55. This suggests that most repositories are reasonably well-structured and easy to maintain. However, a small number of outliers fall below 20, indicating cases where significant maintainability issues may exist.

## Cyclomatic Complexity (CC)

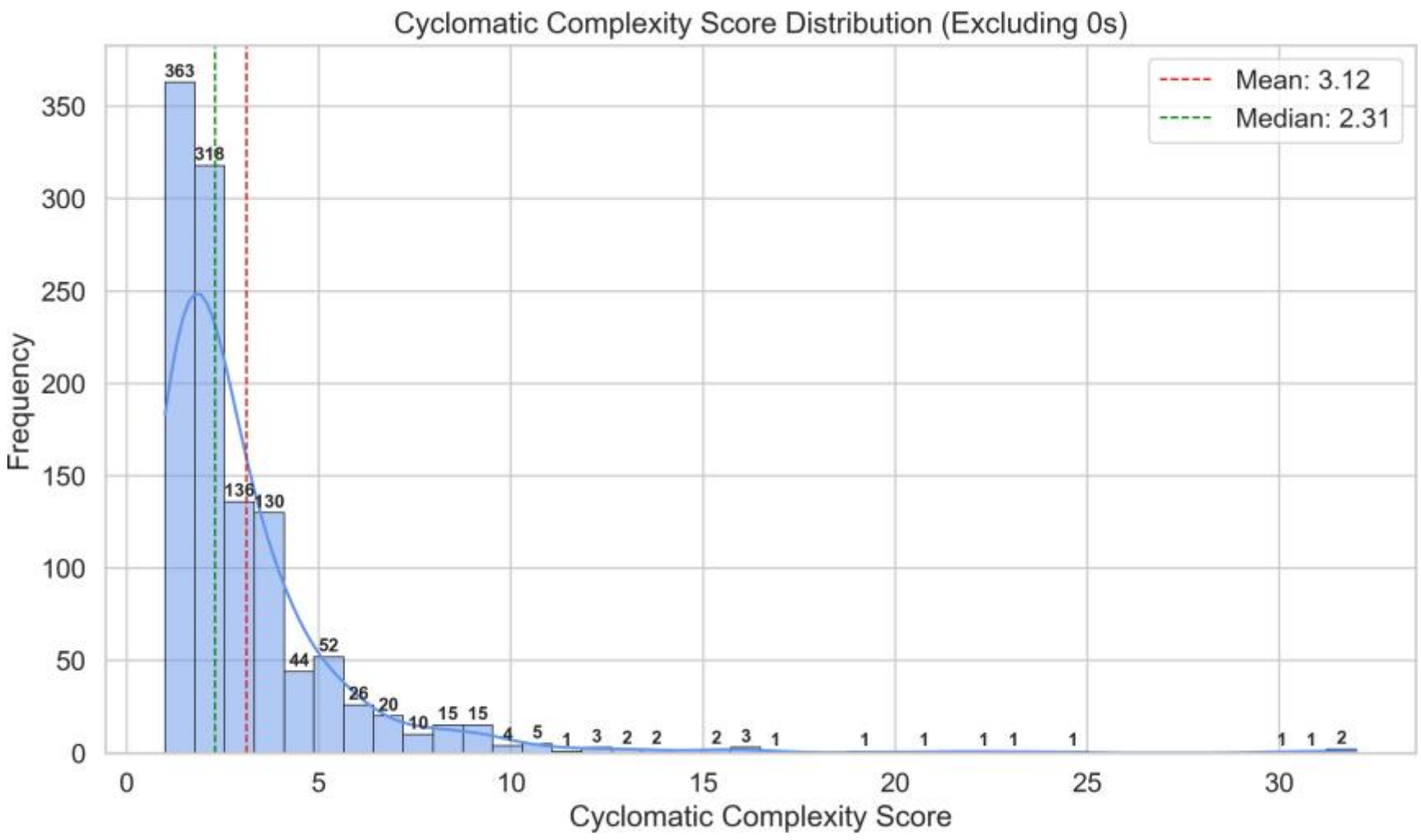

**Fig. 7**: Cyclomatic Complexity Score Distribution

From the histogram in Figure 7, we can see that cyclomatic complexity is generally low across the dataset. The median is 2.31, and the mean is 3.12. It implies that most code blocks are reasonably straightforward and structurally manageable. However, a few repositories have cyclomatic complexity scores exceeding 10, indicating more complex control flow or insufficient function refactoring Watson et al. (1996).

## Code Smells (CS): PyLint Score

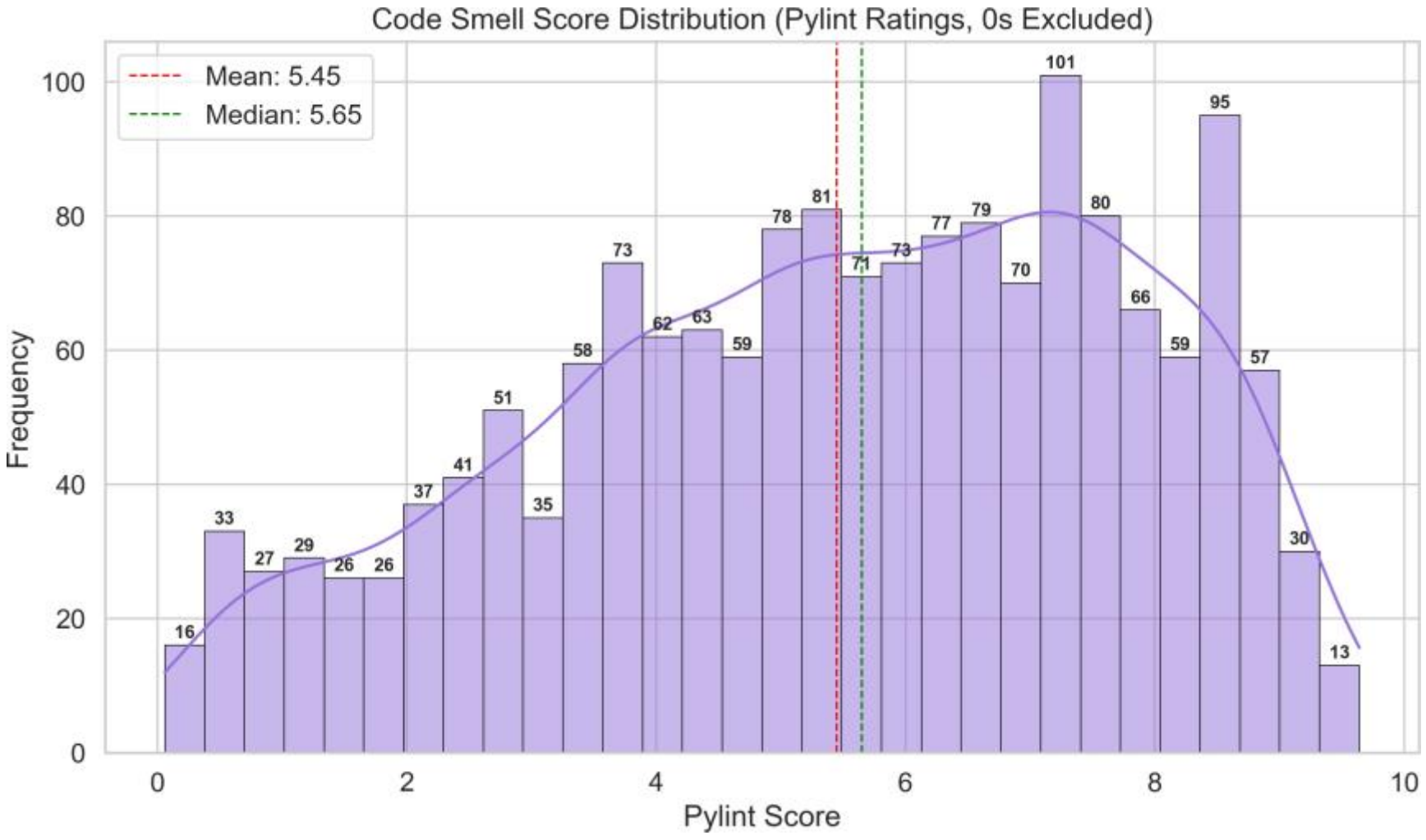

**Fig. 8**: Code Smell Score Distribution

From the histogram in Figure 8, we observe that most repositories score between 5 and 8 on Pylint, with a median of 5.65 and a mean of 5.45. It suggests an average level of code quality, with room for improvement. A few notebooks have very low scores, reflecting fewer potential code issues.

### Comment Density (CoDe)

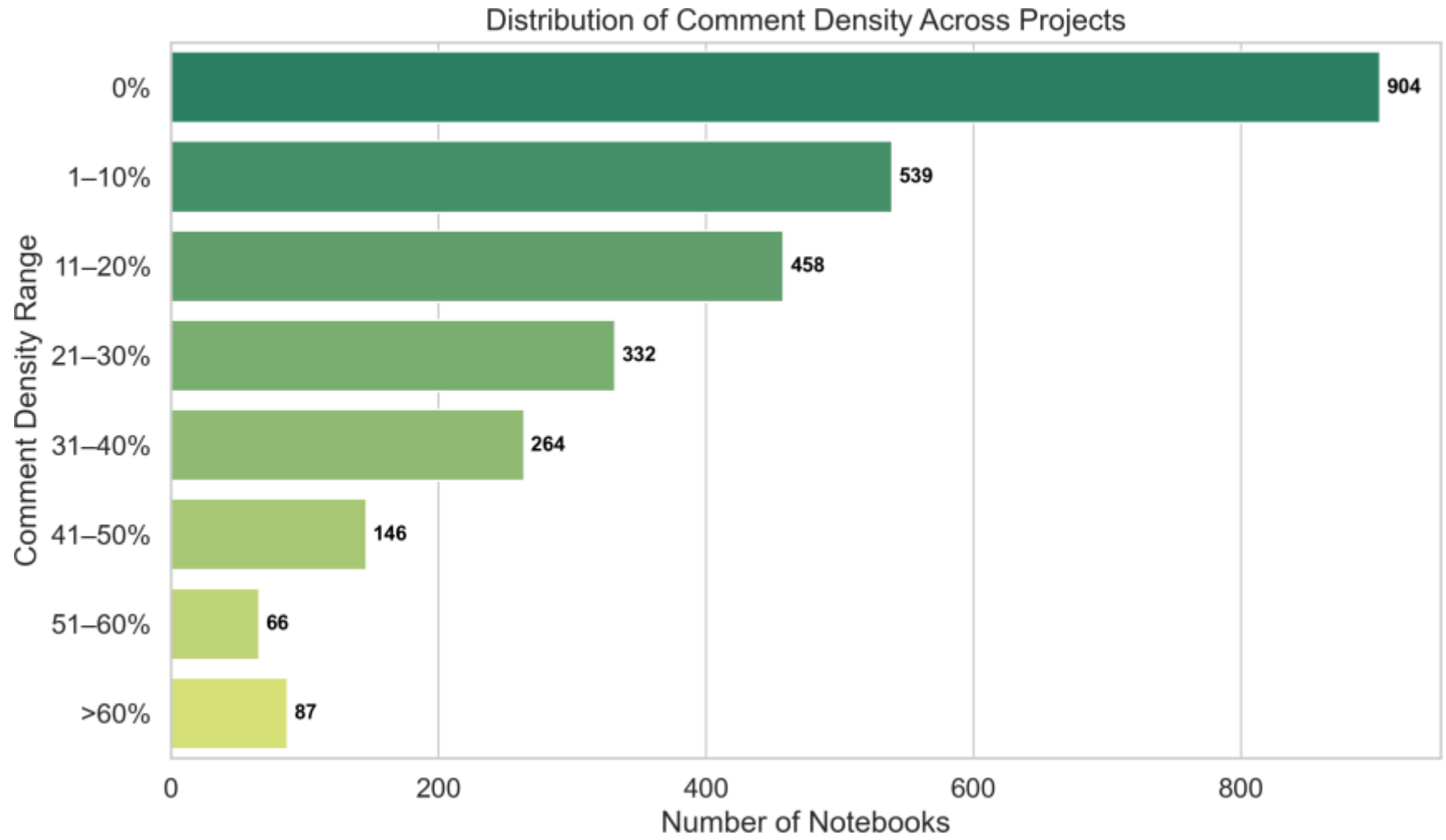

**Fig. 9**: Distribution of Comment Density

The horizontal bar chart in Figure 9 demonstrates a significant number of repositories (Jupyter notebooks or Python files) (904 out of 2,796) have no comments at all. Even though many notebooks contain documentation, the overall trend is towards insufficient documentation. This lack of comments can hinder readability and reuse, especially for external collaborators.

## Code Duplication (CoDu)

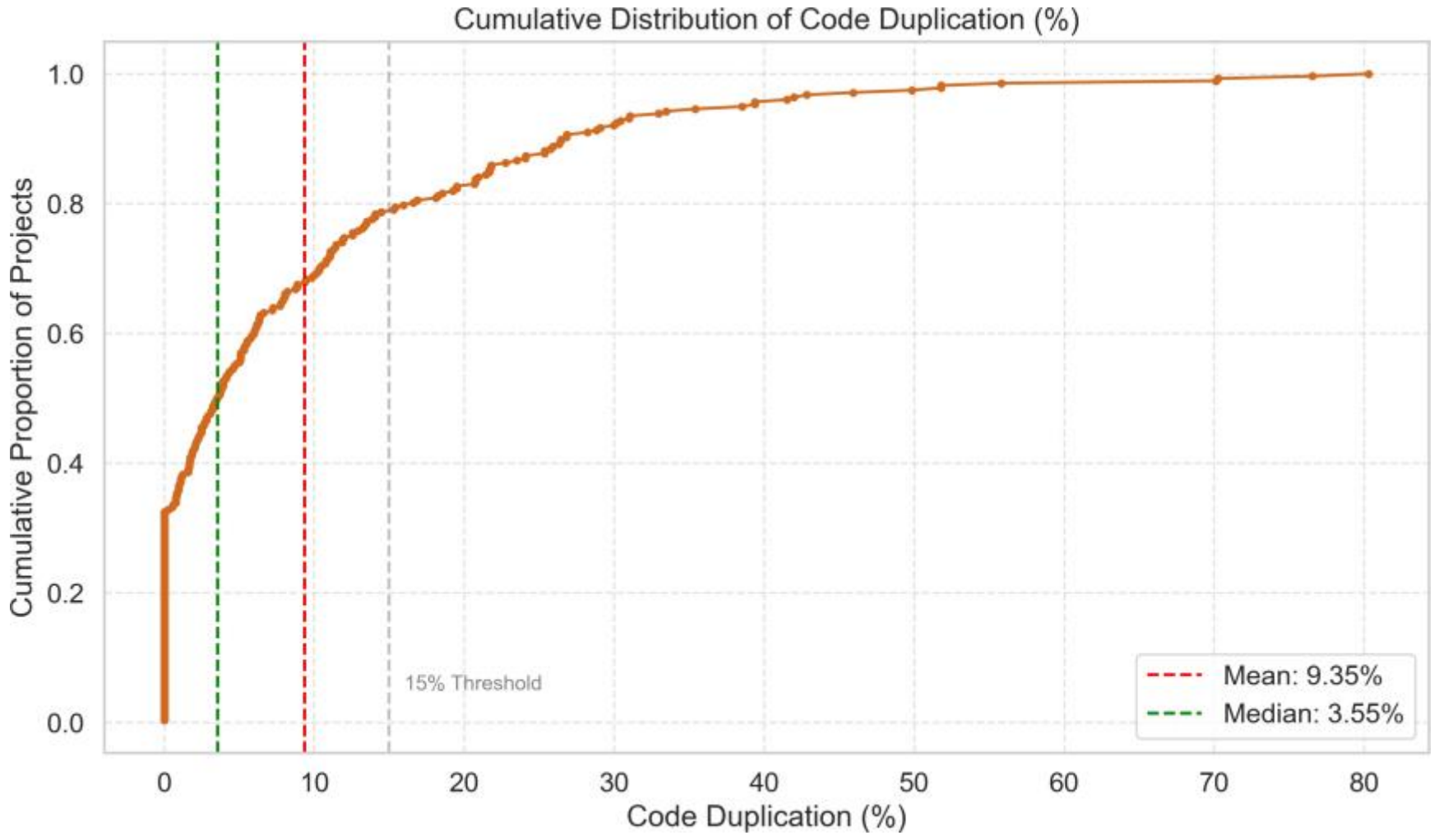

**Fig. 10**: Cumulative Distribution of Code Duplication (%)

### 5.4.2 Improving Software Quality

In this section, we use code (functions) from the selected 10 repositories to provide feedback using the quality control engine to improve code quality with LLMs. We used initial feedback from our quality control engine and combined it with code and a curated prompt to generate higher-quality code. The generated code is then scanned again using the script and sent to the LLM for further generation. We ran this loop for five iterations. Our experimental analysis shows that, on average, after four to five iterations, the metrics stop improving. The prompt is highlighted in Figure 11.

**Explaining Table 9:** We conducted the experiments using three prompting strategies. First, we evaluated our proposed approach using iterative dynamic feedback from the quality control engine. Second, we used a single structured feedback signal without iterative refinement. Third, as an ablation setting, we provided only the prompt and source code to the LLM, without feedback from the quality-control engine.

For the dynamic-feedback approach, we evaluated both OpenAI LLMs (green) and local HuggingFace LLMs. During experiments with the local models, smaller models (e.g., LLaMA, Qwen, and Mistral models below 16B parameters) frequently produced code that could not be executed because of syntax or formatting issues, including the insertion of Markdown code delimiters ("'), inappropriate semicolons in Python code, incorrect indentation, and cases in which the entire codebase was placed inside a comment. Larger local models (e.g., LLaMA, Qwen, and Mistral models above 16B parameters) also exhibited reliability issues. In some cases, they retained only

```
You are an expert in software engineering. Based on the given code, general
    code issues, and specific code issues, generate a better version of the
    code such that the error messages are improved.

ADD DETAILED COMMENTS (**Explain logic and intent**) -> Higher score is
    better
PUT CODE INSIDE MULTIPLE FUNCTIONS FOR MAINTAINABILITY (**Modularize into
    small functions**) Higher score is better
MAKE SURE THERE IS NO DUPLICATE CODE (**Eliminate repetition with
    abstractions**) Lower score is better
MAKE SURE THE CODE IS MAINTAINABLE (**Readable, modular, well-tested design
    **) Higher score is better
LOW CYCLOMATIC COMPLEXITY (**Target Ideal complexity range**) Ideal score
    ranging from 1-10 is better

Given Code:
{code_str}

General Code Issue:
{general_error_message}

Special Code Issue:
{special_error_message}

Only generate code.
DO NOT GENERATE ANY QUOTATION LIKE '''
```

**Fig. 11**: The curated prompt provided to the LLM

a subset of the functions from the input during the first iteration, while subsequent iterations could remove the entire codebase. This produced misleading quality measurements, such as a 100% Maintainability Index and 0% Code Smells for effectively empty output. Therefore, for the experiments reported below, we retained only the first-iteration results for the local LLMs and excluded these models from the other prompting strategies because of their unreliable outputs.

Throughout the following experimental tables, CS refers to Code Smells, LoC to Lines of Code, CS/LoC to Code Smells normalized by Lines of Code, MI to Maintainability Index, CoDu to Code Duplication, CoDe to Comment Density, and CC to Cyclomatic Complexity. The arrows indicate the preferred direction of improvement, with ↑ denoting higher-is-better and ↓ denoting lower-is-better.

In Table 9, our dynamic feedback setup improves several structural quality attributes over the baseline, particularly code duplication and code-smell-related metrics. We compared the baseline for the 10 repositories (CS = 8.30, MI = 73.96, CoDu = 33.07, CoDe = 45.37, LoC = 146, CC = 0.0) with all LLM-based variants. We find that code duplication is consistently eliminated (CoDu = 0.0) and that several models reduce the number of code smells, indicating that the refactored code is generally less redundant and structurally cleaner. However, improvements are not uniform across all metrics. In particular, some models generate substantially larger codebases and achieve lower maintainability index values than the baseline. These observations highlight the multi-objective nature of software quality improvement, where gains in one quality attribute may introduce trade-offs in others.

Among the evaluated models, GPT-4o-mini achieves a strong balance, with substantially fewer code smells (CS = 6.06) and the highest maintainability index (MI =

**Table 9**: Comparison of the Results from our Proposed System with other Methodologies

| Method | LLM | CS ↓ | MI ↑ | CoDu ↓ | CoDe ↑ | LoC | CC |
|---|---|---|---|---|---|---|---|
| Baseline | (No LLM Applied) | 8.30 | 73.96 | 33.07 | 45.37 | 146 | 0.0 |
| LLM Prompting (Unstructured) | GPT 40-mini | 7.46 | 53.37 | 0.0 | 1.21 | 171 | 1.4 |
| | GPT 5-nano | 8.11 | 50.54 | 0.0 | 10.33 | 419 | 4.2 |
| | GPT 4o | 6.41 | 55.13 | 0.0 | 2.12 | 173 | 1.33 |
| LLM + Structured Feedback (Single Iteration) | GPT 4o-mini | 8.59 | 76.33 | 0.0 | 10.22 | 226 | 1.07 |
| | GPT 5-nano | 8.13 | 57.60 | 0.0 | 10.35 | 426 | 2.42 |
| | GPT 4o | 6.71 | 48.63 | 0.0 | 0.0 | 182 | 1.21 |
| LLM + Structured Feedback (Iterative) | GPT 4o-mini | 6.06 | 79.21 | 0.0 | 29.52 | 165 | 1.1 |
| | GPT 5-nano | 8.36 | 65.04 | 0.0 | 8.93 | 333 | 2.0 |
| | GPT 4o | 6.62 | 64.14 | 0.0 | 5.74 | 173 | 1.18 |
| | Mistral 22B | 4.25 | 73.15 | 0.0 | 6.12 | 127 | 1.0 |
| | Qwen 30B-Instruct | 6.4 | 72.05 | 0.0 | 7.61 | 188 | 1.44 |
| | CodeLLaMa 34B-Inst. | ~ | 100.0 | 0.0 | 0 | 81 | 0.0 |
| | CodeLLaMa 70B | ~ | 100.0 | 0.0 | 0.0 | 0 | 0.0 |

*Note:* CS = Code Smells; MI = Maintainability Index; CoDu = Code Duplication; CoDe = Comment Density; LoC = Lines of Code; CC = Cyclomatic Complexity. ↑ indicates that higher values are preferred, whereas ↓ indicates that lower values are preferred.

79.21) while keeping cyclomatic complexity modest (CC = 1.1). Mistral 22B exhibits the lowest code-smell score overall (CS = 4.25), combined with a high maintainability index (MI = 73.15) and relatively compact code (LoC = 127). However, the results from the open-source LLMs are less reliable, as some models tend to remove a significant portion of the original code. In contrast, GPT-5-nano generates considerably longer code (LoC = 333 and 419 in the dynamic and prompt-only configurations, respectively), primarily because it produces extensive inline documentation and multi-line comments for many functions, as illustrated in Figure 13. Despite the increase in code size, the resulting comment density remains relatively low. A closer inspection of the generated code revealed frequent use of try–catch blocks and additional conditional logic, which increase the number of executable lines and therefore reduce the proportion of comments relative to the total code size.

We compared with different feedback-based strategies and concluded that our proposed dynamic feedback shows superior performance compared to single or prompt-only feedback. However, when we moved from dynamic to single or prompt-only feedback, we observed a consistent degradation in MI (e.g., GPT-4o-mini: 79.21 → 76.33 → 53.37) and a minor reduction in code smells. We also observe that comment density often drops sharply and becomes very low for some settings (e.g., CoDe values around 1–10 for most single/prompt-only runs). This observation suggests that a single refinement step or purely prompt-based guidance is insufficient to guide the models toward generating code that follows the quality model we defined.

**Explaining Table 10:** The results in Table 10 summarize how code quality evolves over multiple refinement steps when using GPT-5-nano in our dynamic feedback loop. Although it is unclear whether the Code Smells (CS) metric is improving, we use CS/LoC to quantify code smells per line of code. Therefore, in column 3 (CS/LoC), we observe that it decreases from 5% to 2.01% at step 4. Moreover, at Step 0, the baseline code already has a relatively high Maintainability Index (MI = 73.99), but it exhibits substantial code duplication (CoDu = 33.07) and a relatively compact size (LoC = 146), with a high comment density (CoDe = 45.37). After the first refinement step, the model eliminates duplication (CoDu = 0), which is a clear structural improvement, but this comes with trade-offs: the code becomes noticeably longer (LoC = 271), MI drops (67.45), and comment density falls sharply (CoDe = 7.74) (reasoning explained in answering RQ1). In other words, the model restructures the code to eliminate redundant fragments, thereby improving code quality.

**Table 10**: Code Quality Improvement Over Multiple Steps

| | CS ↓ | CS/LoC ↓ | MI ↑ | CoDu ↓ | CoDe ↑ | LoC | CC |
|---|---|---|---|---|---|---|---|
| **Step 0** | 7.34 | 5% | 73.99 | 33.07 | 45.37 | 146 | 0 |
| **Step 1** | 7.78 | 2.8% | 67.45 | 0 | 7.74 | 271 | 1.85 |
| **Step 2** | 7.72 | 2.3% | 63.12 | 0 | 10.32 | 329 | 2.11 |
| **Step 3** | 7.92 | 2.3% | 64.04 | 0 | 18.85 | 340 | 2.05 |
| **Step 4** | 8.07 | 2.01% | 60.30 | 0 | 16.95 | 386 | 2.22 |

As refinement progresses from Step 1 to Step 4, this tension persists. Additional steps further increase the code size (LoC grows from 271 to 386) and slightly raise cyclomatic complexity (from 1.85 to 2.22), which remains low but reflects the introduction of more branches or modularization. Comment density recovers slightly after Step 1, reaching 18.85 at Step 3. However, it dropped slightly again at Step 4, suggesting that the model incrementally reintroduces or restructures comments as it refines the code. Meanwhile, MI gradually decreases across steps (from 73.99 at baseline to 60.30 at Step 4), and code smells (CS) increase marginally, illustrating an important point: improving one aspect of quality (e.g., eliminating duplication and adding structure) may impact others, due to increased code length and new control flow. Overall, the table shows that multi-step refinement with GPT-5-nano is effective at addressing specific issues, such as duplication. Still, it also highlights the inherently multi-objective nature of code quality, where gains in one metric can come at the cost of another unless those trade-offs are explicitly controlled.

**Explaining Table 11:** In this part, we will try to analyze the effect of fixing one metric while the others are ignored. To conduct this experiment, we updated our quality control engine to provide feedback only on a single metric while evaluating the code on all other metrics. The rows in table 11 suggest that the metric in the first row is provided to the LLM to update, and when the LLM generates the new code, we use it to evaluate with all the metrics. While the ideal condition is that the provided metric would improve (compared to baseline), and others would decrease, we found mixed results. We observe that MI decreased relative to the baseline when we asked the LLM to improve MI alone. On the other hand, when we asked to improve Cyclomatic Complexity, it only increased to 2.68; however, when we asked to improve Comment Density, CC increased to 2.875. This also suggests that these metrics are interconnected: when one metric changes, others may improve or deteriorate. Therefore, we need to prioritize which metrics to improve and determine which combinations of metrics increase them.

**Table 11**: Empirical pairwise interactions among maintainability metrics

| | CS | MI | CoDu | CoDe | CC |
|---|---|---|---|---|---|
| **CS** | 7.44 | 49.58 | 0.0 | 8.75 | 2.60 |
| **MI** | 8.2 | 51.86 | 0.0 | 6.95 | 2.66 |
| **CoDu** | 7.06 | 49.09 | 0.0 | 13.23 | 2.74 |
| **CoDe** | 7.7 | 51.54 | 0.0 | 8.24 | 2.875 |
| **CC** | 8.22 | 42.21 | 0.0 | 6.87 | 2.68 |
| **Baseline** | 8.36 | 65.04 | 0.0 | 8.93 | 2.0 |

```
1
2  X_train, X_test, Y_train, Y_test = train_test_split(X, Y, test_size = 0.33,
       random_state = 42)
3  X_train.shape, X_test.shape, Y_train.shape, Y_test.shape
4
5  # ## Training the model
6
7  model=RandomForestClassifier()
8
9  model.fit(X_train, Y_train)
10
11 # ## Evaluating the model via Accuracy
12 X_train_prediction=model.predict(X_train)
13 train_accuracy=accuracy_score(X_train_prediction,Y_train)
14 print('Accuracy of training data :',train_accuracy)
15
16 X_test_prediction=model.predict(X_test)
17 test_accuracy=accuracy_score(X_test_prediction,Y_test)
18 print('Accuracy of testing data :',test_accuracy)
```

**Fig. 12**: A code to train and evaluate a RandomForest Classifier Model (Before Updating)

### 5.4.3 Qualitative Analysis on Improving Software Quality

In this experiment, we analyze and present a qualitative study of how our system improves code quality compared to the original one. Figure 12 shows an example of a sample code we picked from the repository we used to answer RQ2. Figure 13 shows the corresponding LLM-refactored version of the code. However, compared to Figure 12, Figure 13 shows a significant improvement in code modularization and the addition of additional comments to make the code easier to understand. As a result, the code will be more maintainable and easier to understand.

Compared to the developer-written baseline, which trains a 'RandomForestClassifier' (Figure 12) inline with a single comment header and hard-coded variables, the code produced by our system (Figure 13) substantially improves several core quality attributes. First, it adds a rich, structured docstring that clearly explains the logic and intent of 'train_model', including parameter semantics and return types, yielding

```python
def train_model(X: pd.DataFrame, y: pd.Series,
                random_state: int,
                n_estimators: int,
                max_depth: Optional[int]) -> RandomForestClassifier:
    """
    Build and train a RandomForestClassifier on the given data.
    Parameters
    ----------
    X : pd.DataFrame
        Feature matrix used for training.
    y : pd.Series
        Target labels corresponding to the rows in X.
    random_state : int
        Seed for the random number generator to ensure reproducibility.
    n_estimators : int
        Number of trees in the forest.
    max_depth : Optional[int]
        Maximum depth of each tree in the forest. If None, no maximum is
    enforced.
    Returns
    -------
    RandomForestClassifier
        Trained random forest model.
    try:
        model = build_model(
            nestimators=nestimators,
            random_state=random_state,
            max_depth=max depth,
        )
        model.fit(X, y)
        return model
    except Exception as e:
        # You can replace this with proper logging if desired
        raise RuntimeError(
            f"Failed to train RandomForestClassifier: {e}"
        ) from e
```

**Fig. 13**: A modularized and well-commented clean code to train and evaluate a RandomForest Classifier Model (After Updating)

a much higher score on "ADD DETAILED COMMENTS." Second, by encapsulating the logic in 'train_model' and delegating construction to 'build model', it "PUTS CODE INSIDE MULTIPLE FUNCTIONS FOR MAINTAINABILITY," promoting reuse, testability, and more precise separation of concerns. This functional abstraction also proactively addresses "MAKE SURE THERE ARE NO DUPLICATE CODE," since future training scenarios can reuse the same functions instead of copy–pasting classifier instantiation and fitting logic. Together with explicit parameters, type hints, and the elimination of global dependencies (e.g., 'X_train', 'Y_train'), these design choices significantly enhance "MAKE SURE THE CODE IS MAINTAINABLE" by improving readability, modularity, and ease of evolution. Finally, while both snippets demonstrate low cyclomatic complexity in isolation, the modular structure of the generated version provides a better foundation for keeping complexity in the desirable medium range as additional behavior (e.g., logging, validation, or tuning) is added, thereby supporting long-term control-flow manageability.

# 6 Discussion

The results demonstrate how a lifecycle-aware Quality Model can support both systematic assessment and LLM-based improvement of Tier 1 research software. In the following, we discuss the main findings by answering the three research questions defined in Section 3.4.

**RQ1. Designing a Lifecycle-Aware Quality Model:** A useful Quality Model for Tier 1 research software needs to balance conceptual completeness with practical measurability. While the conceptual model captures a broad range of quality concerns, only a subset can currently be assessed automatically or semi-automatically. Moreover, the results show that software quality cannot be represented reliably by a single metric, since software that performs well on maintainability or complexity may still have weaknesses in documentation, duplication, dependencies, or other aspects. Lifecycle awareness provides additional context by relating quality concerns to the stages in which they are most relevant. Therefore, a practical Quality Model for Tier 1 research software should be multidimensional, operationalizable, and sensitive to lifecycle context.

**RQ2. Using Quality Metrics for LLM-Based Refinement:** Quality metrics can serve not only as assessment measures but also as structured feedback for LLM-based refinement. The comparison of feedback strategies indicates that iterative structured feedback provides more effective guidance than unstructured prompting or a single feedback step for some models. The qualitative analysis further shows that this guidance can result in concrete improvements in modularization, documentation, and separation of responsibilities. However, the single-metric experiments show that quality properties are interconnected, and targeting one metric can also affect others. Structured feedback should therefore consider the overall quality profile and be updated through repeated assessment rather than treating individual metrics as independent optimization targets.

**RQ3. Effectiveness of the Feedback-Driven Framework:** The framework can improve specific quality attributes, particularly code duplication and code smells, but the improvements are not uniform across all metrics. Additional refinement iterations can improve some properties while negatively affecting maintainability, code size, comment density, or other measures, confirming that software quality improvement is a multi-objective problem. Effectiveness also depends on the capability of the LLM, as some local models produced invalid or incomplete code while still obtaining favorable static-analysis scores. The framework is consequently most effective as a guided quality improvement mechanism in which the overall quality profile, interactions among metrics, and model reliability are considered together.

**Limitations of our System** The current study also presents several opportunities for further extension and validation. While the conceptual Quality Model includes 25 metrics, 6 were operationalized automatically or semi-automatically in the present implementation, reflecting the availability of reliable tooling and infrastructure for notebook-centric environments. Similarly, although the model encompasses Maintainability, Security, FAIRness, Functional Suitability, and Sustainability, the LLM-based refinement experiments focus primarily on maintainability-related metrics, providing a foundation for future evaluation across the remaining quality dimensions. Finally, the Decision Module currently reports improvements and regressions across individual metrics rather than explicitly optimizing among competing quality objectives.

## 7 Conclusion and Future Work

In this work, we addressed the challenge of improving the quality of Tier-1 research software, with a particular focus on notebook-centric workflows. Research software is often developed under time pressure by researchers without formal software-engineering training, which can lead to quality issues that hinder reproducibility, reuse, and long-term sustainability. Although static analysis tools can identify such issues, their outputs may be difficult for non-expert programmers to interpret and translate into actionable improvements. To address this gap, we proposed a lifecycle-aware quality control framework that combines a quantitative Quality Model with an LLM-driven feedback loop. The Quality Model identifies and operationalizes quality metrics relevant to Tier-1 research software and transforms the resulting diagnostics into structured, lifecycle-aware feedback that guides iterative LLM-based refinement.

Our experimental evaluation demonstrates that the proposed feedback-driven approach can improve several quality attributes compared with the original research code and alternative prompting strategies. In particular, the results show improvements in code duplication and structural quality, while the qualitative case study illustrates how metric-driven feedback can guide research code toward more modular and maintainable implementations. However, improvements are not uniform across all metrics: optimizing one quality attribute can negatively affect others, such as code size, complexity, or documentation-related measures. These findings highlight the inherently multi-objective nature of software quality improvement and the importance of considering the overall quality profile rather than optimizing individual metrics in isolation.

The results demonstrate the potential of combining quantitative quality assessment with iterative LLM-based refinement to support researchers in improving Tier-1 research software. At the same time, the absence of test suites in many of the evaluated repositories limited systematic validation of the functional correctness of generated code. Future work will therefore extend the framework with functional validation, broader quality-metric coverage, and more advanced multi-objective optimization strategies. Further investigation of interactions among quality metrics and evaluation with researchers in real-world workflows will also be important for assessing the framework's effectiveness across different research software contexts.